# Crowd behaves as excitable media during Mexican wave

Mexican wave, or La Ola, first widely broadcasted during the 1986 World Cup held in Mexico, is a human wave moving along the stands of stadiums as one section of spectators stands up, arms lifting, then sits down as the next section does the same (Fig.1a). Here we use variants of models originally developed for the description of excitable media to demonstrate that this collective human behaviour can be quantitatively interpreted by methods of statistical physics. Adequate modelling of reactions to triggering attempts provides a deeper insight into the mechanisms by which a crowd can be stimulated to execute a particular pattern of behaviour and represents a possible tool of control during events involving excited groups of people.

Using video recordings we have analysed 14 waves in stadiums with above 50.000 people: the wave has a typical velocity in the range of 12*m/s* (20 seats/s), a width of about 6-12*m* (~15 seats) and more frequently rolls in the clockwise direction. It is generated by the simultaneous standing up of not more than a few dozens of people and subsequently expands over the entire tribune acquiring its stable, close to linear shape (see http://angel.elte.hu/wave dedicated to the present work, offering further data and interactive simulations).

The relative simplicity of the Mexican wave allows us to develop a quantitative treatment of this kind of collective behaviour by building and simulating models accurately reproducing and predicting the details of the associated human wave. We show here that the well-established approaches to the theoretical interpretation of excitable media[1-3] - originally created for describing such processes as forest fires or wave propagation in heart tissue  - can readily be generalized to include human social behaviour.

We have developed two mathematical simulation models, a minimal and a more detailed one to demonstrate the robustness of our approach. In analogy with models of excitable media, in both versions people are regarded as excitable units: they can be activated by an external stimulus (a distance and direction-wise weighted concentration of nearby active people exceeding a threshold value *c*).  Once activated, each unit follows the same set of internal rules to pass through the active (standing and waving) and refractory (passive) phases before returning to its original, resting (excitable) state. While the simpler version distinguishes three states only (excitable/active/passive) and accounts for variations in the individual behaviour by means of transition probabilities between the states, the elaborate version takes into account an actual, deterministic activity pattern in more detail. The two versions of the model we considered differ in the way stochasticity, i.e., differences and fluctuations regarding the above behavioural patterns are represented (for details see http://angel.elte.hu/wave).

Next, we employed these models to get an insight into the conditions for triggering a wave. Figure 1b shows the evolution of a wave provoked by the simultaneous excitation (standing up) of a small group of units (people). Using parameters deduced from video recordings for the sizes and characteristic times of the phenomenon (interaction radius, reaction/activation times and probabilities) we have been able to reproduce the above described observations concerning the size/form/velocity and stability of the wave.  Fig.1c displays the probability of generating a wave when a small group of varying size tries to trigger it under different excitation threshold values.

Our results clearly demonstrate that the dependence of the eventual occurrence of a wave on the number of initiators is a rather sharply changing function, i.e., triggering a Mexican wave requires a critical mass. The present approach is expected to have implications for the treatment of situations where influencing the behaviour of a crowd is desirable. In particular, in the context of violent street incidents associated with demonstrations or sport events, it is essential to know under what conditions groups can gain control over the crowd and how fast and in which form this perturbation/transition can spread.

**I. Farkas\*, D. Helbing+ and T. Vicsek\***

\* Department of Biological Physics, Eötvös University Budapest, H-1117 Hungary
+ Institute for Economics and Traffic, Dresden University of Technology, G-01062 Germany
*e-mail: vicsek@angel.elte.hu*

1. Wiener, N & Rosenblueth, A., *Arch. Inst. Cardiol. Mexico*. **16**, 205-265 (1946)
2. Greenberg, J.M. & Hastings, S.P. *SIAM J. of Appl. Math*. **34**, 515-523 (1978)
3. Bub, G., Shrier, A. & Glass, L. *Phys. Rev. Lett*. **88**, 058101 (2002)

___

Figure caption:

**Figure 1**. Photo and simulations of the Mexican wave. *Model:* If the weighted concentration of active people within a radius of $R$ around a person is above the threshold of the person $c_i$ (randomly chosen from [$c-\Delta c$, $c+\Delta c$]) then the person is activated. Weights decrease exponentially with distance and change linearly with the *cosine* of the direction so that people on the left of a person have an influence $w_0$ times as strong as those on the right. The direction of the wave's motion is determined by this anisotropy due to spontaneous symmetry breaking at the early stages resulting from anticipation and the anisotropy in perception since the majority of people are right handed. 1a) Photo of a Mexican wave. 1b) Snapshots of the $n$-state model, where, after activation, a person deterministically goes through $n_a$ active states (stages of standing up) and $n_r$ refractory states. The wave is shown at 0.5$s$, 2$s$, and 15$s$ after the triggering event on a tribune with 80 rows of seats. Brighter shades correspond to higher level of activity. Parameters are $n_a=n_r=5$, $c=0.25$, $\Delta c=0.05$, $R=3$ and $w_0=0.5$. 1c) $P(N,c)$, the ratio of successful triggering events, as a function of the number of people $N$ in the group trying to induce a wave and the average threshold $c$. Parameters are as above, and each point represents the average of 128 simulations.

a)

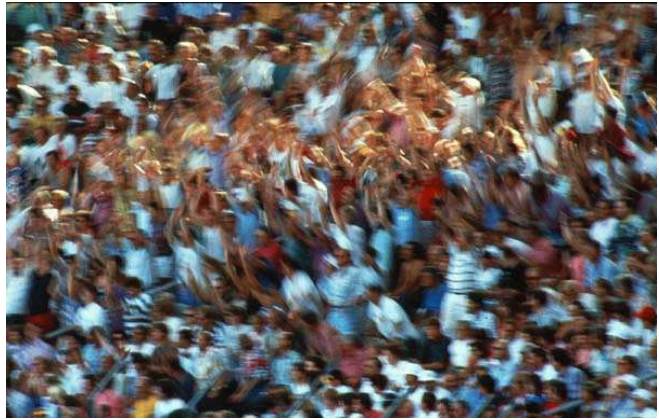

b)

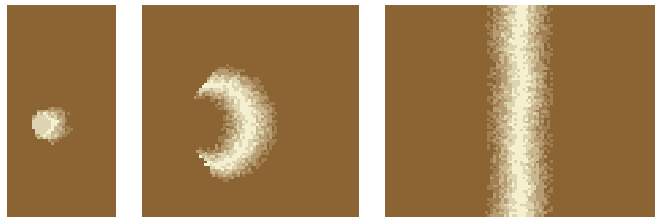

c)

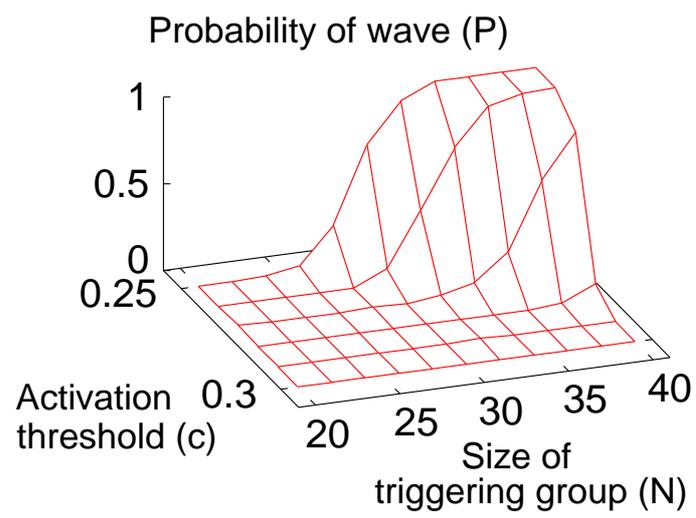

FIGURE 1.